\documentclass[aps,pre,twocolumn,groupedaddress,10pt,showpacs,superscriptaddress]{revtex4-1}
\usepackage{amsmath,amssymb}
\usepackage{graphicx,color}

\newcommand{\dif}{\mathrm{d}}
\newcommand{\fdif}{\operatorname{\delta}}
\newcommand{\Fdif}[2]{\frac{\fdif\!#1}{\fdif\!#2}}
\newcommand{\pdif}[2]{\frac{\partial#1}{\partial#2}}
\newcommand{\Laplace}{\boldsymbol{\triangle}}
\newcommand{\Nabla}{\vec{\nabla}}

\begin{document}

\title{Stability of liquid crystalline phases in the phase-field-crystal model} 

\author{Cristian Vasile Achim}
\affiliation{Department of Applied Physics, Alto University School of Science, P.O. Box 11000, FI-00076 Aalto, Finland}
\affiliation{Institut f{\"u}r Theoretische Physik II, Weiche Materie,
Heinrich-Heine-Universit{\"a}t D{\"u}sseldorf, D-40225 D{\"u}sseldorf, Germany}

\author{Raphael Wittkowski}\author{Hartmut L{\"o}wen}
\affiliation{Institut f{\"u}r Theoretische Physik II, Weiche Materie,
Heinrich-Heine-Universit{\"a}t D{\"u}sseldorf, D-40225 D{\"u}sseldorf, Germany}

\date{\today}

\begin{abstract}
The phase-field-crystal model for liquid crystals is solved numerically
in two spatial dimensions. This model is formulated with three position-dependent order parameters,
namely the reduced translational density, the local nematic order parameter, and the mean
local direction of the orientations. The equilibrium free-energy functional
involves local powers of the order parameters up to fourth order, gradients of the order
parameters up to fourth order, and different couplings between the order parameters.
The stable phases of the equilibrium free-energy functional
are calculated for various coupling parameters. Among the stable
liquid-crystalline states are the isotropic, nematic, columnar, smectic A, and plastic
crystalline phases.
The plastic crystals can have triangular, square, and honeycomb lattices and exhibit orientational 
patterns with a complex topology involving a sublattice with topological defects.
Phase diagrams were obtained by numerical minimization of
the free-energy functional.
Their main features are qualitatively in line with much simpler one-mode approximations
for the order parameters.
\end{abstract}


\maketitle


\section{\label{sec:introduction}Introduction}
Crystallization and melting processes (see e.g.\ \cite{Singh1991,Loewen1994a}) can be efficiently modelled
within the so-called phase-field-crystal (PFC) model \cite{ElderKHG2002,ElderG2004,Emmerich2009} which
is basically a Landau-type theory
with a conserved position-dependent order parameter which is gradient-expanded up to fourth order.
Stable periodic oscillations in the order are interpreted as
crystalline density fields. The PFC model was applied to various situations including fluid-crystal 
interfaces \cite{YuLV2009,JaatinenAEAN2009}, crystal growth into a supercooled liquid \cite{vanTeeffelenBVL2009},
grain boundaries \cite{TegzeGTPJANP2009,McKennaGV2009}, and  the Asaro-Tiller-Grinfeld instability
\cite{AsaroT1972,Grinfeld1986,HuangE2008,WuV2009}. Later on, the PFC model was systematically derived from
density functional theory \cite{Evans1979,Singh1991,Loewen1994a,ElderPBSG2007,vanTeeffelenBVL2009} which provides a microscopic
approach to freezing and melting in equilibrium \cite{RamakrishnanY1979,RosenfeldSLT1997,RothELK2002,HansenGoosM2009}
using a gradient expansion  in terms of density modulations \cite{LoewenBW1989,OhnesorgeLW1991,Lutsko2006,ElderPBSG2007}.
Moreover, dynamical density functional theory for the non-equilibrium dynamics of Brownian systems
\cite{MariniMT1999,ArcherE2004,EspanolL2009} can be used to derive the dynamics of the PFC model \cite{vanTeeffelenBVL2009}.

Recently, the PFC model was generalized to anisotropic particles
by using the appropriate generalizations of density functional theory to orientational degrees of freedom 
\cite{PoniewierskiH1988,GrafL1999,HansenGoosM2009} and to orientational Brownian dynamics
\cite{RexWL2007,WensinkL2008}.
This was done both in two \cite{Loewen2010} and three \cite{WittkowskiLB2010} spatial dimensions \cite{FussnoteMkhonta}.
The resulting PFC models are valid for liquid crystals
composed of apolar particles and describe also liquid-crystalline phases with a uniaxial orientation
distribution. The theory is formulated in terms of three order parameter fields, namely the reduced translational 
density, the local nematic order parameter, and the orientational direction. It includes gradients up
to fourth order in the reduced translational density field and up to second order in the remaining
orientational order parameters. While the original PFC model has two free parameters,
the liquid-crystalline PFC model in two dimensions \cite{Loewen2010} has five independent
couplings. This widely opens the parameter space for the occurrence of several
liquid-crystalline phases including nematic, columnar, smectic A, plastic crystalline, and orientationally ordered crystalline
phases \cite{Frenkel1991,BolhuisF1997}. However, no numerical calculation of the PFC model has been presented for the stability
of these different liquid-crystalline phases, yet.

In this paper, we start from the PFC model for liquid crystals in two spatial dimensions
proposed in reference \cite{Loewen2010} and determine the stable liquid-crystalline
phases numerically for special coupling-parameter combinations. Among the stable
liquid-crystalline phases are the isotropic, nematic, stripe, columnar, smectic A, and plastic triangular
crystalline phase. The latter exhibit complex orientational patterns with a sublattice of topological defects.
For stronger translational-orientational couplings, a plastic crystal with fourfold square symmetry
is getting stable. This does not occur for the traditional PFC model \cite{ElderKHG2002,ElderG2004,FussnoteOettel}
but can be induced for other Ginzburg-Landau functionals \cite{SakaguchiB1997}.
Also a plastic honeycomb crystal was observed.
We found second-order phase transitions for the isotropic-nematic phase transition and phase transitions of first order for all other phase transformations.

The paper is organized as follows: Sec.\ \ref{sec:PFC} is addressed to the PFC model for liquid crystals. 
In \ref{subsec:FEF}, we recall the free-energy functional for two-dimensional liquid crystals and introduce a simpler form of this functional, which is obtained by an alternative choice of the order-parameter fields.
Two different methods for the minimization of this functional are presented in \ref{subsec:NM}.
Then, in Sec.\ \ref{sec:NR}, results for these two methods are shown.
We finally conclude in Sec.\ \ref{sec:conclusions}.

\section{\label{sec:PFC}Phase-field-crystal model}
\subsection{\label{subsec:FEF}Free-energy functional}
\begin{figure*}[ht]
\centering
\begin{tabular}{ccccc}
\multicolumn{2}{c}{\textbf{isotropic}} 
&\phantom{aaaai} &\multicolumn{2}{c}{\textbf{nematic}}\\
\includegraphics[height=0.2\linewidth]{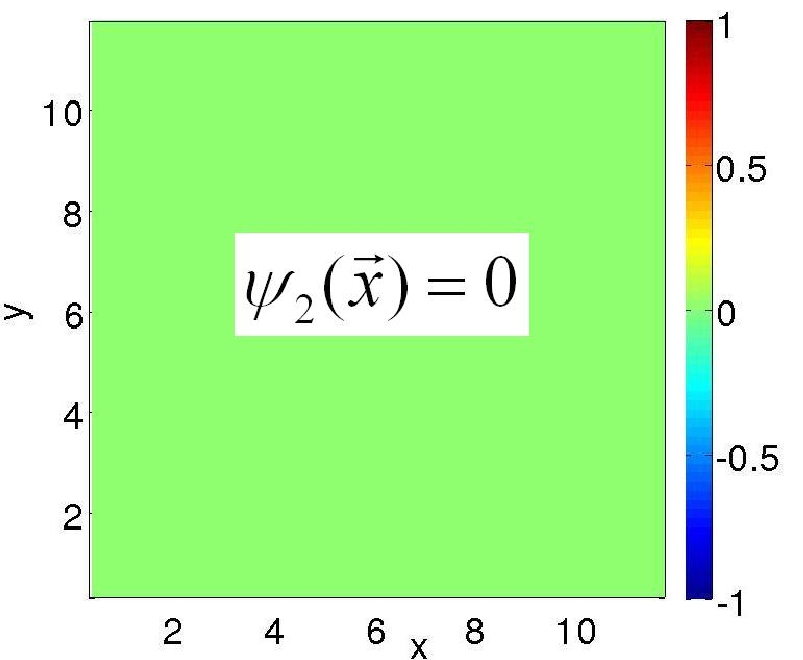}
&\includegraphics[height=0.2\linewidth]{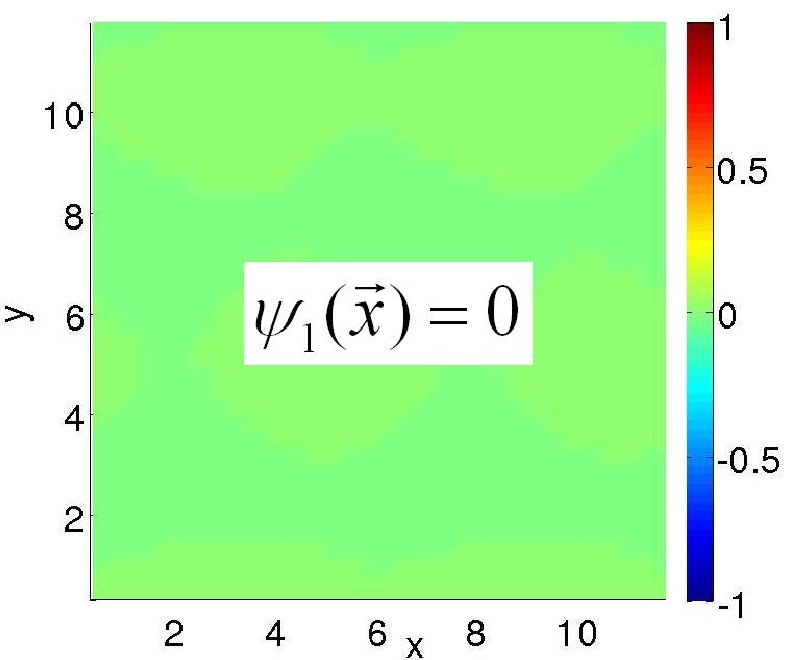}
&&\includegraphics[height=0.2\linewidth]{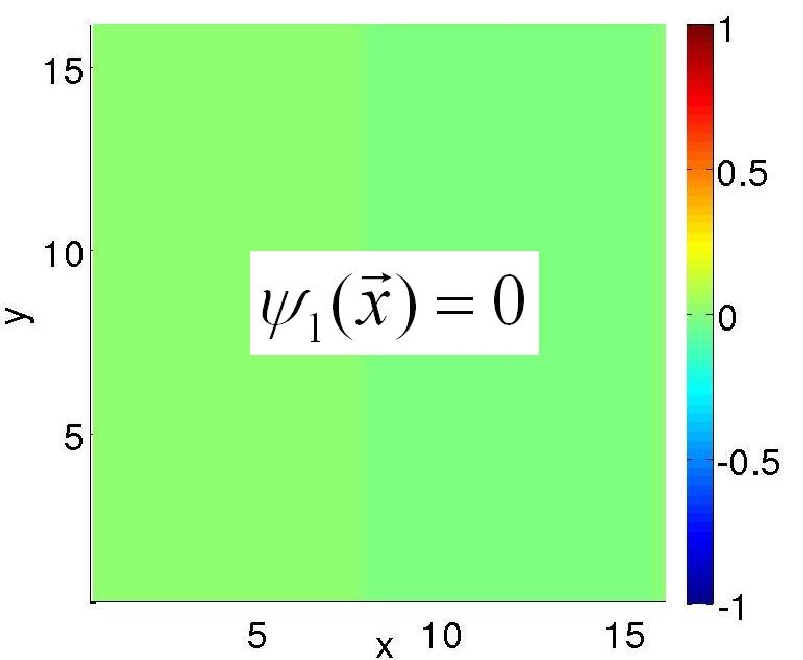}
&\includegraphics[height=0.2\linewidth]{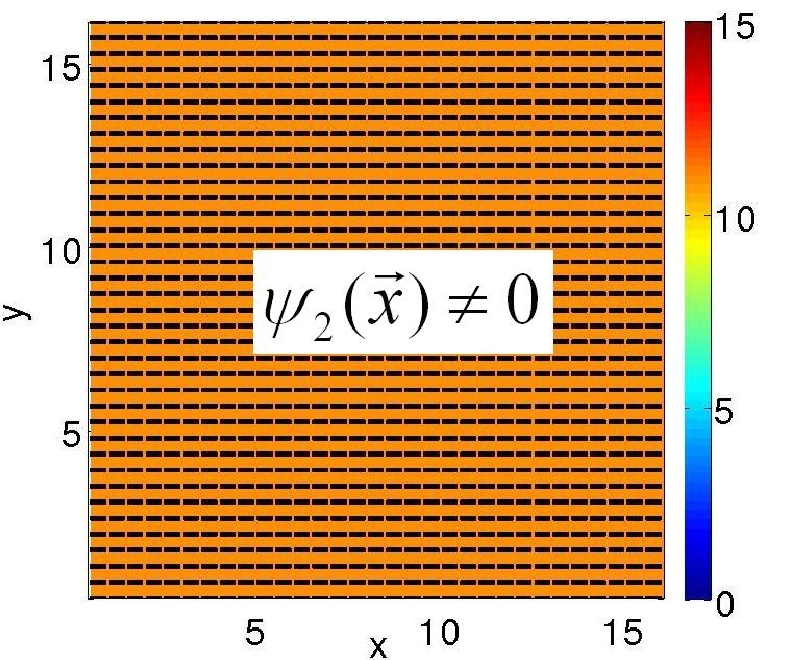}\\
\multicolumn{2}{c}{\textbf{stripes}} 
&&\multicolumn{2}{c}{\textbf{columnar/smectic A}}\\
\includegraphics[height=0.2\linewidth]{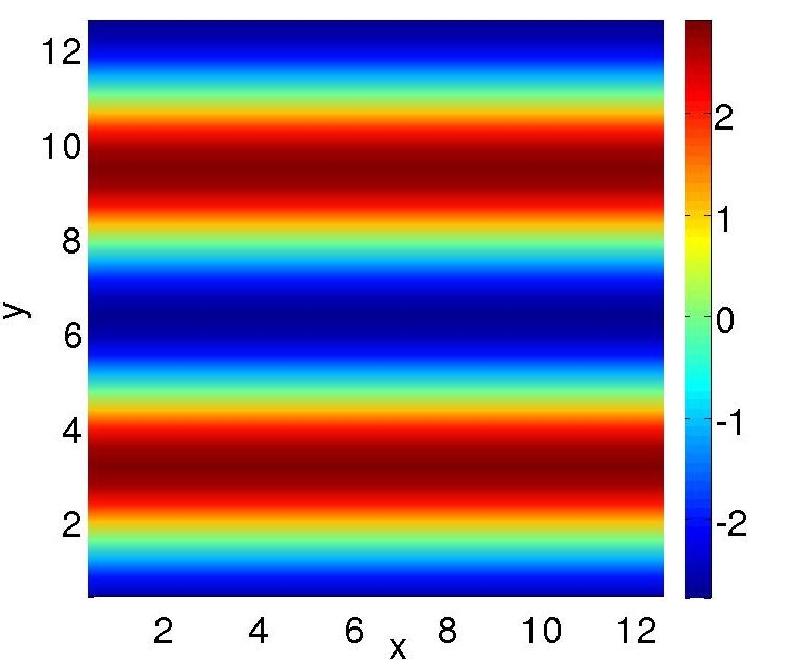}
&\includegraphics[height=0.2\linewidth]{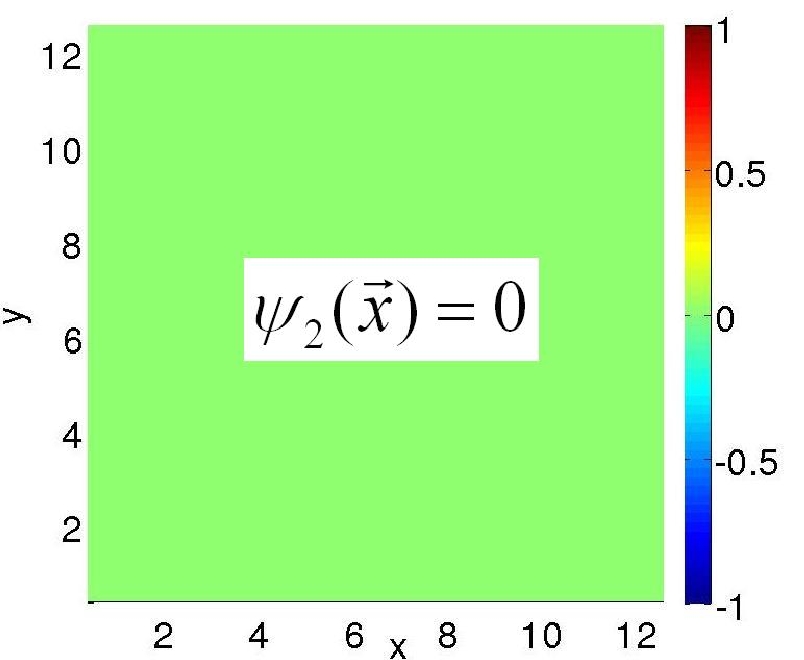}
&&\includegraphics[height=0.2\linewidth]{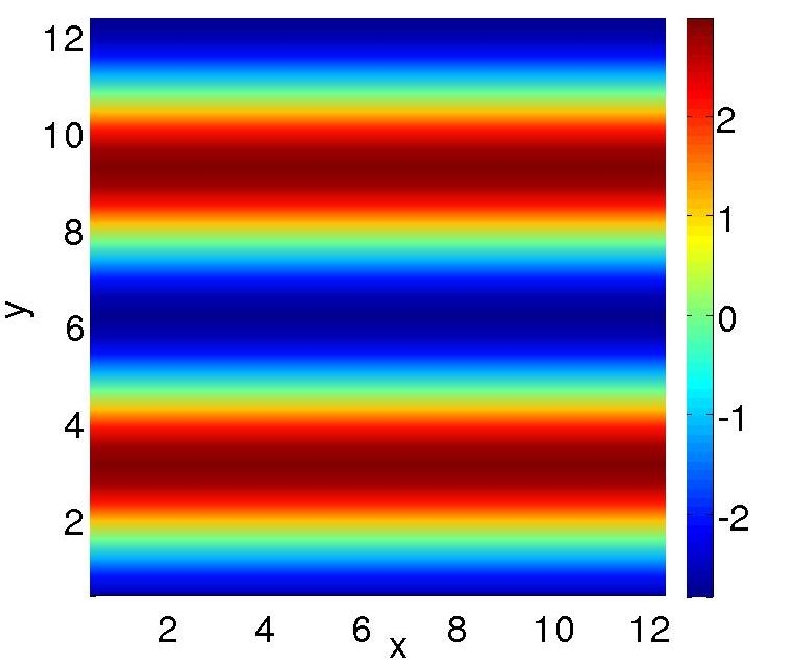}
&\includegraphics[height=0.2\linewidth]{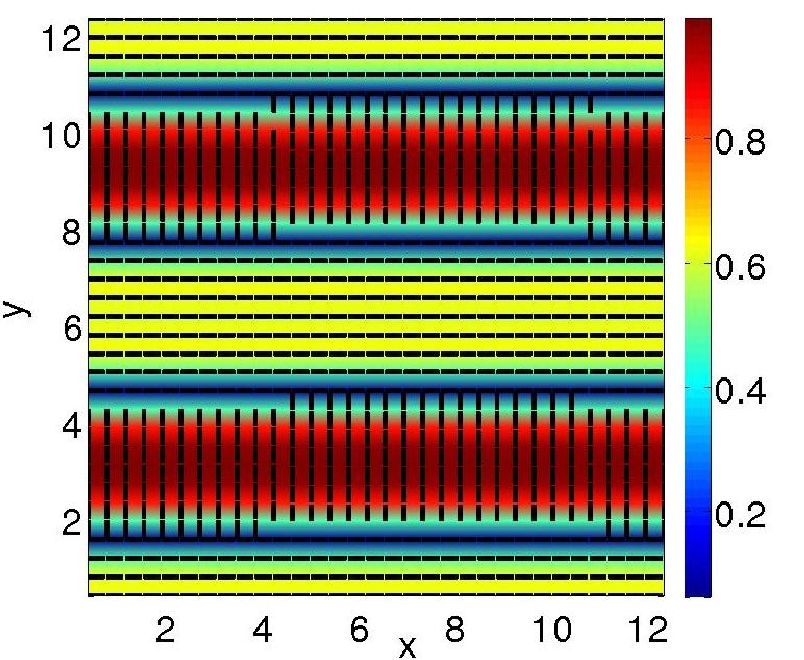}\\
\multicolumn{2}{c}{\textbf{plastic triangular crystal 1}} 
&&\multicolumn{2}{c}{\textbf{plastic triangular crystal 2}}\\
\includegraphics[height=0.2\linewidth]{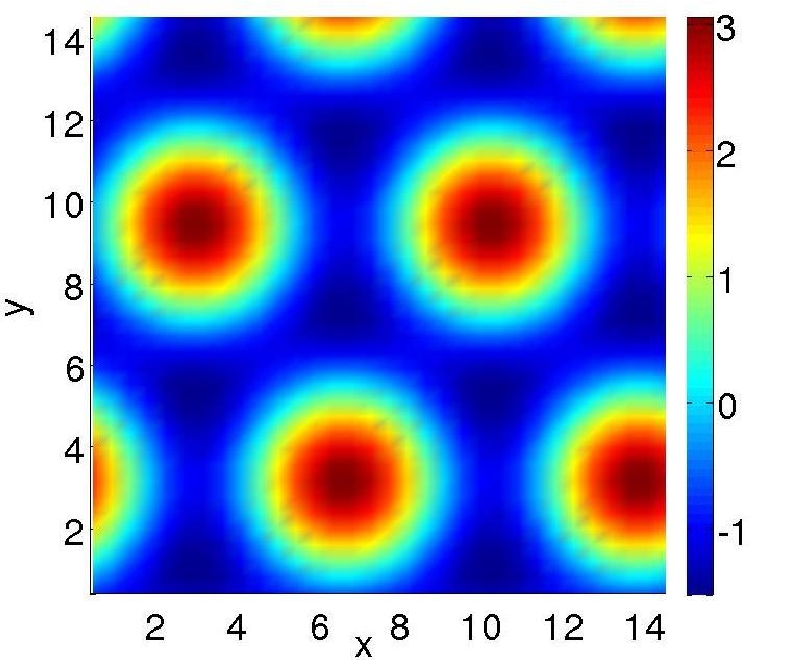}
&\includegraphics[height=0.2\linewidth]{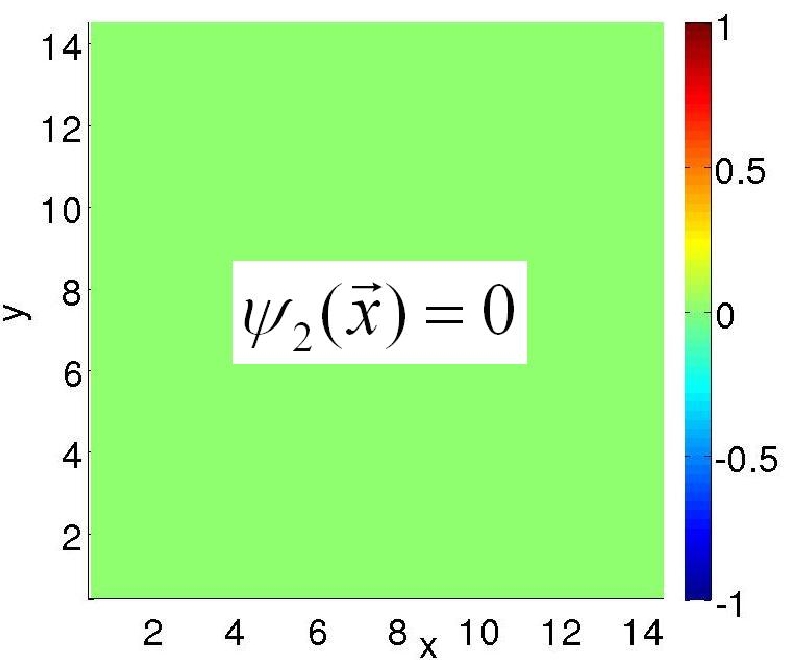}
&&\includegraphics[height=0.2\linewidth]{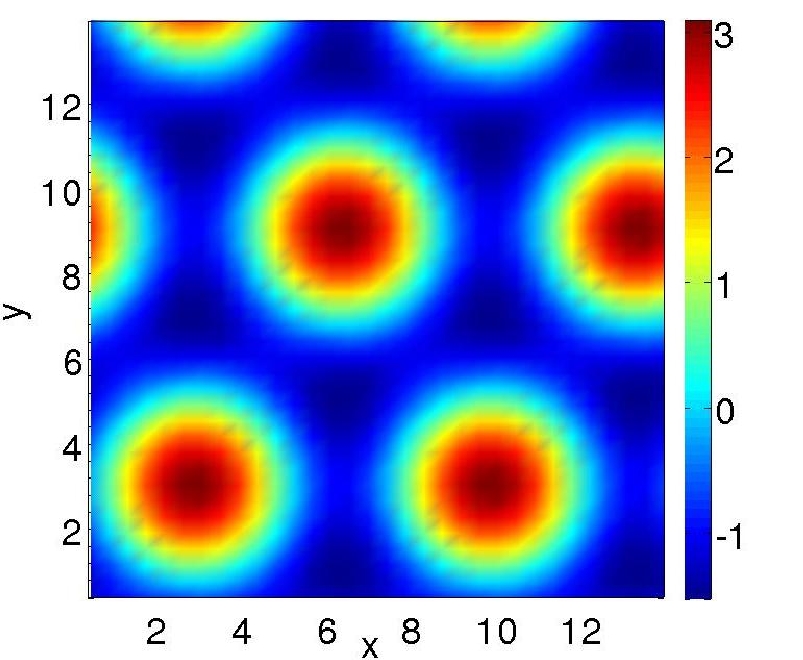}
&\includegraphics[height=0.2\linewidth]{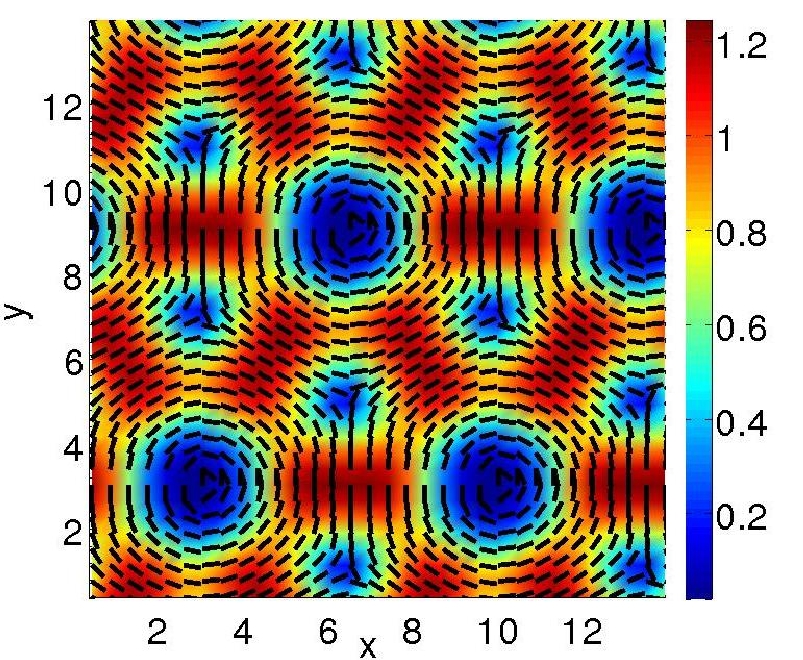}\\
\multicolumn{2}{c}{\textbf{plastic honeycomb crystal}} 
&&\multicolumn{2}{c}{\textbf{plastic square crystal}}\\
\includegraphics[height=0.2\linewidth]{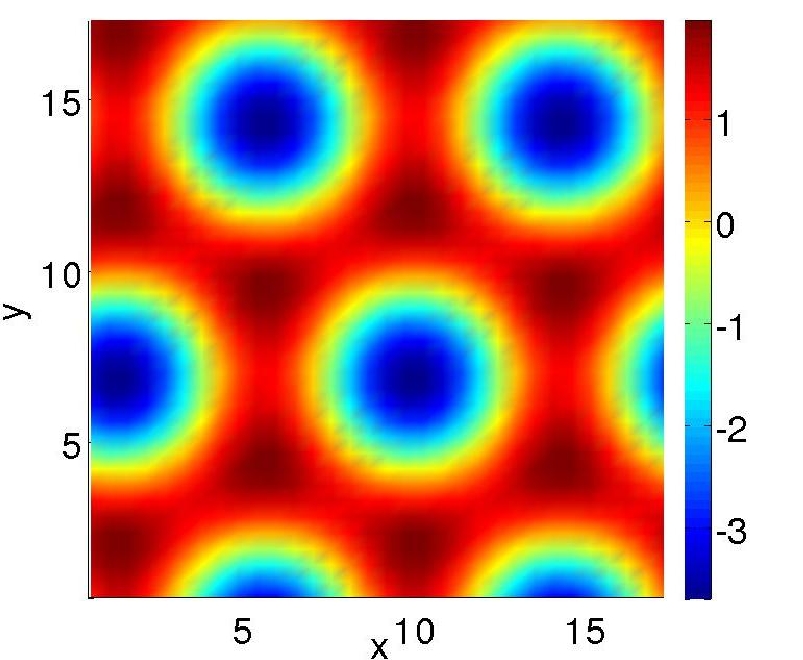}
&\includegraphics[height=0.2\linewidth]{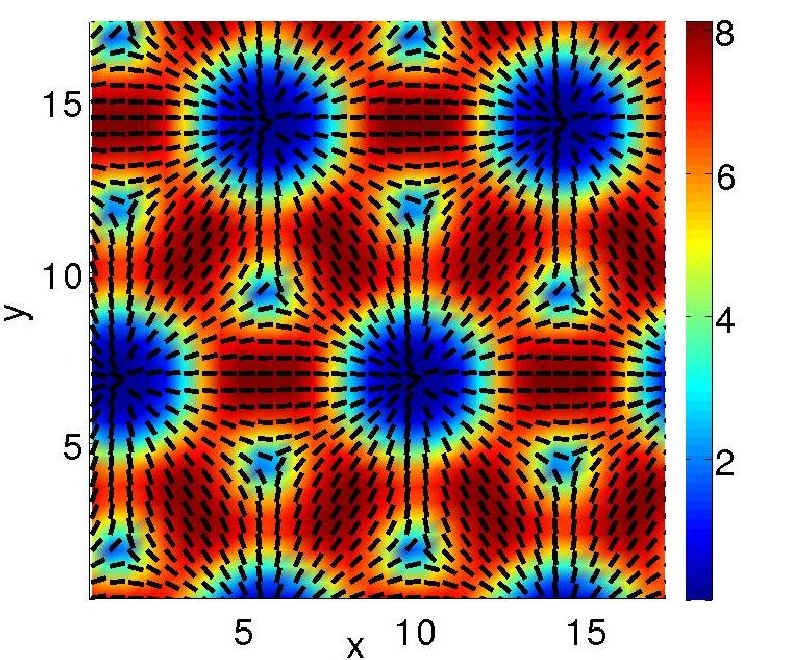}
&&\includegraphics[height=0.2\linewidth]{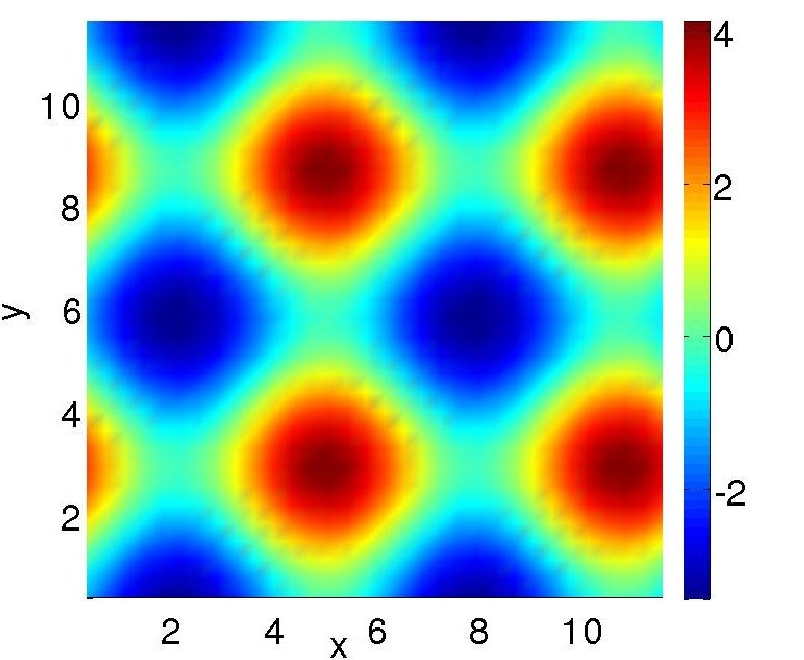}
&\includegraphics[height=0.2\linewidth]{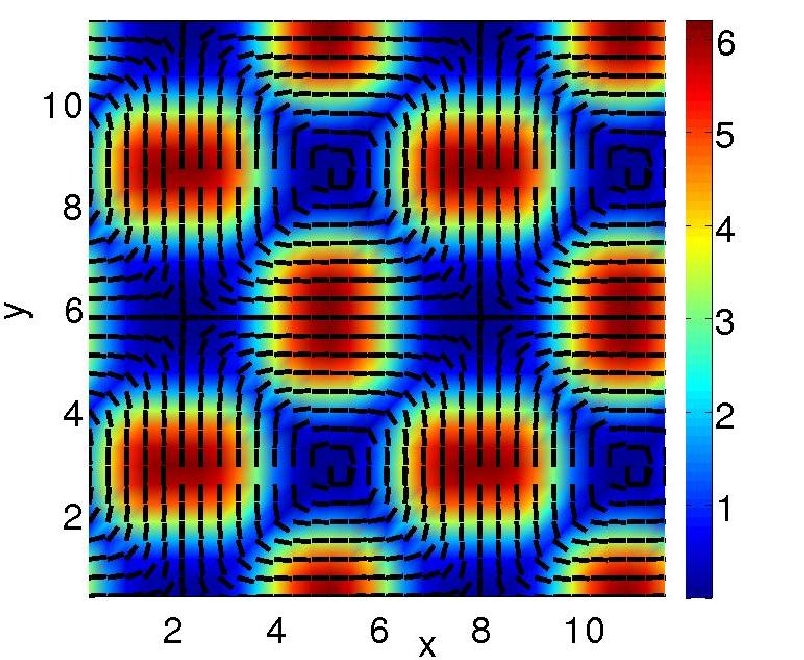}\\
$\psi_{1}(\vec{x})$
&$\psi_{2}(\vec{x})$
&&$\psi_{1}(\vec{x})$
&$\psi_{2}(\vec{x})$\\
\end{tabular}
\caption{\label{fig:PA}Stable liquid crystalline phases. The contour plots show the order-parameter fields $\psi_{1}(\vec{x})$ and $\psi_{2}(\vec{x})$ for the isotropic and nematic phase, the stripe phase and columnar/smectic A phase, two plastic triangular crystals with different orientational ordering, and a plastic honeycomb crystal as well as a plastic square crystal. The black lines in the plots of the second and fourth column represent the director field $\hat{u}_{0}(\vec{x})$. In the cases with $\psi_{2}(\vec{x})=0$ it is not shown, because it is not defined.
The parameters are $B_{\mathrm{x}}=3.5$, $E=1$, and $F=0$ for the stripe phase and the plastic triangular crystal 1 and $B_{\mathrm{x}}=3.5$, $E=0.1$, and $F=1$ for all other phases.}%
\end{figure*}
\begin{figure}[ht]
\centering
\begin{tabular}{c}
\textbf{plastic triangular crystal 2}\\
\includegraphics[width=0.6\linewidth]{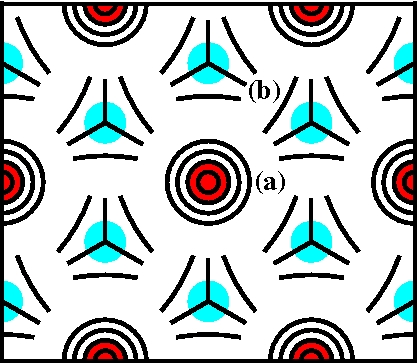}\\[3mm]
\textbf{plastic honeycomb crystal}\\
\includegraphics[width=0.6\linewidth]{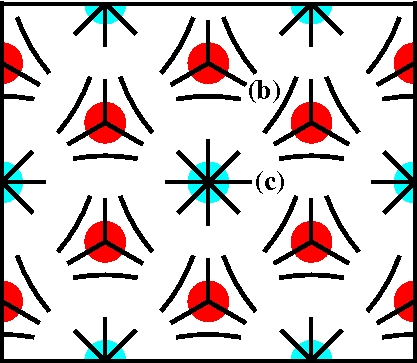}\\[3mm]
\textbf{plastic square crystal}\\
\includegraphics[width=0.6\linewidth]{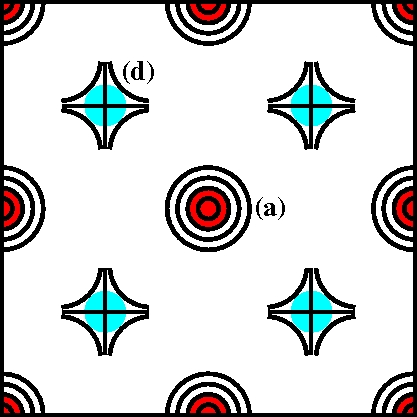}\\
\end{tabular}
\caption{\label{fig:D}Topological defects in three different plastic liquid crystals (schematic). The defects coincide with the maxima (red discs) and minima (cyan discs) of the translational density field $\psi_{1}(\vec{x})$. The symbols in the plots represent the following defects: vortices (a) with the topological winding number $m=1$, disclinations (b) with $m=-1/2$, sources/sinks (c) with $m=1$, and hyperbolic points (d) with $m=-1$.}%
\end{figure}
\begin{figure}[ht]
\centering
\includegraphics[width=0.485\linewidth]{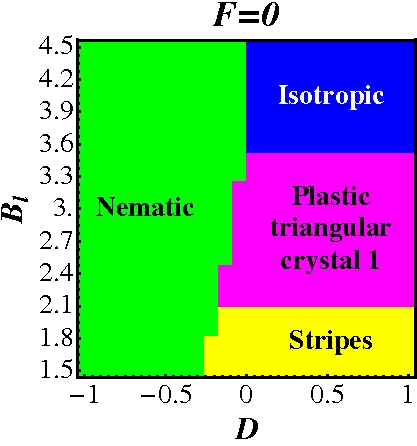}
\,
\includegraphics[width=0.485\linewidth]{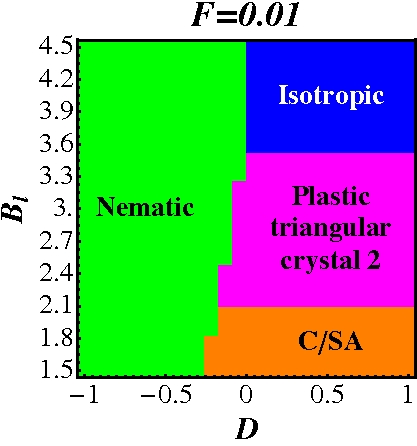}\\
\includegraphics[width=0.485\linewidth]{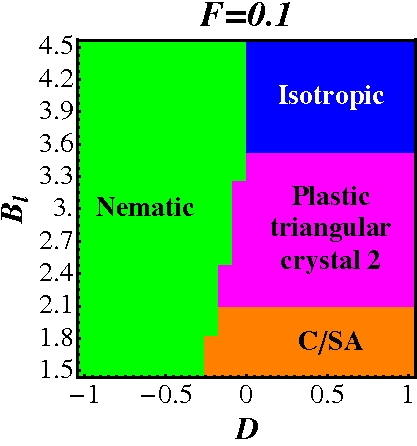}
\,
\includegraphics[width=0.485\linewidth]{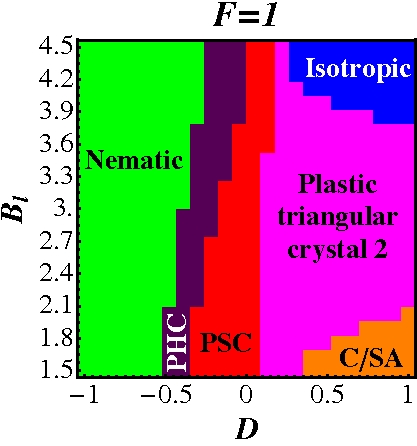}\\
\caption{\label{fig:PDA}Phase diagrams calculated by full numerical minimization for the parameters $B_{\mathrm{x}}=3.5$ and $E=0.1$. The relevant liquid crystalline phases are: isotropic (blue), nematic (green), stripes (yellow), columnar/smectic A (C/SA, orange), plastic triangular crystals (magenta), plastic honeycomb crystal (PHC, dark purple), and plastic square crystal (PSC, red).}%
\end{figure}
\begin{figure}[ht]
\centering
\begin{tabular}{cc}
\multicolumn{2}{c}{\textbf{isotropic}}\\
\includegraphics[height=0.4\linewidth]{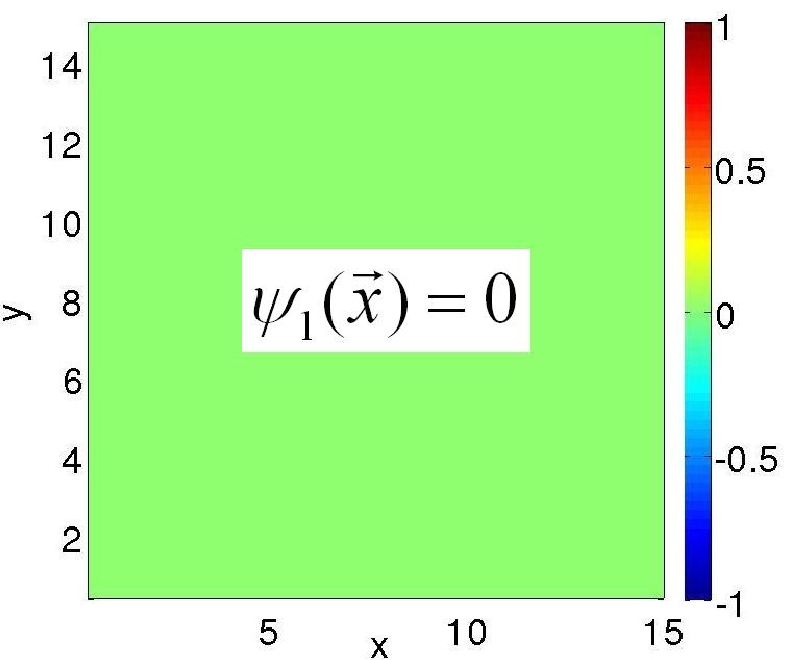}
&\includegraphics[height=0.4\linewidth]{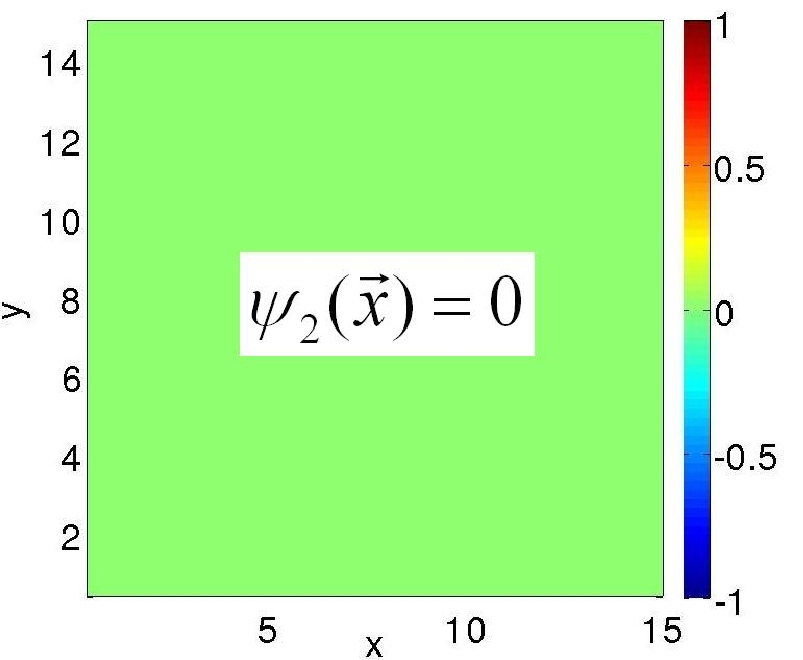}\\
\multicolumn{2}{c}{\textbf{nematic}}\\
\includegraphics[height=0.4\linewidth]{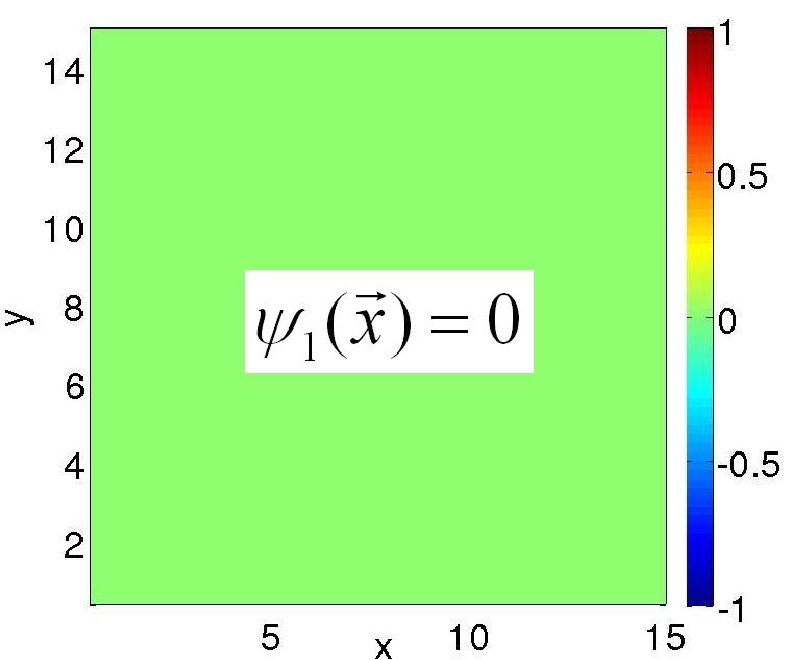}
&\includegraphics[height=0.4\linewidth]{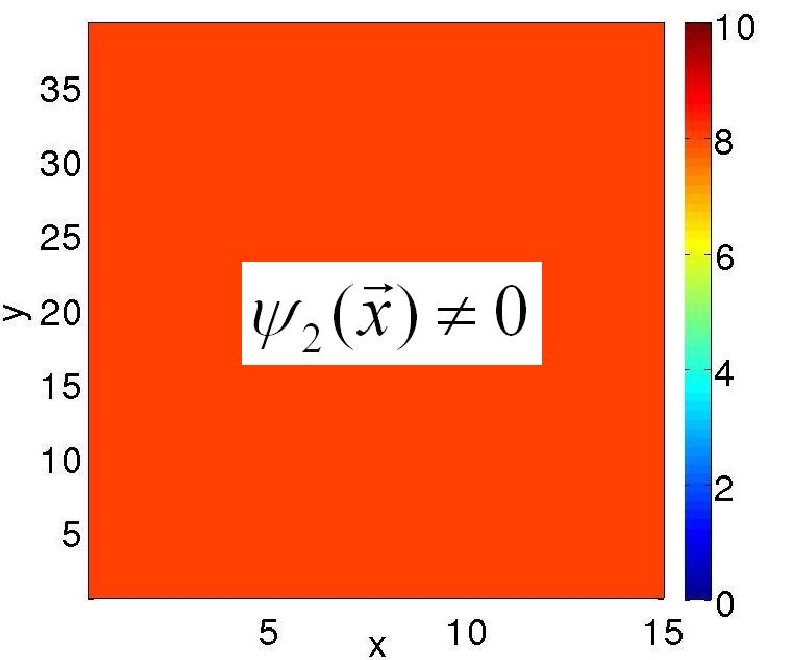}\\
\multicolumn{2}{c}{\textbf{stripes}}\\
\includegraphics[height=0.4\linewidth]{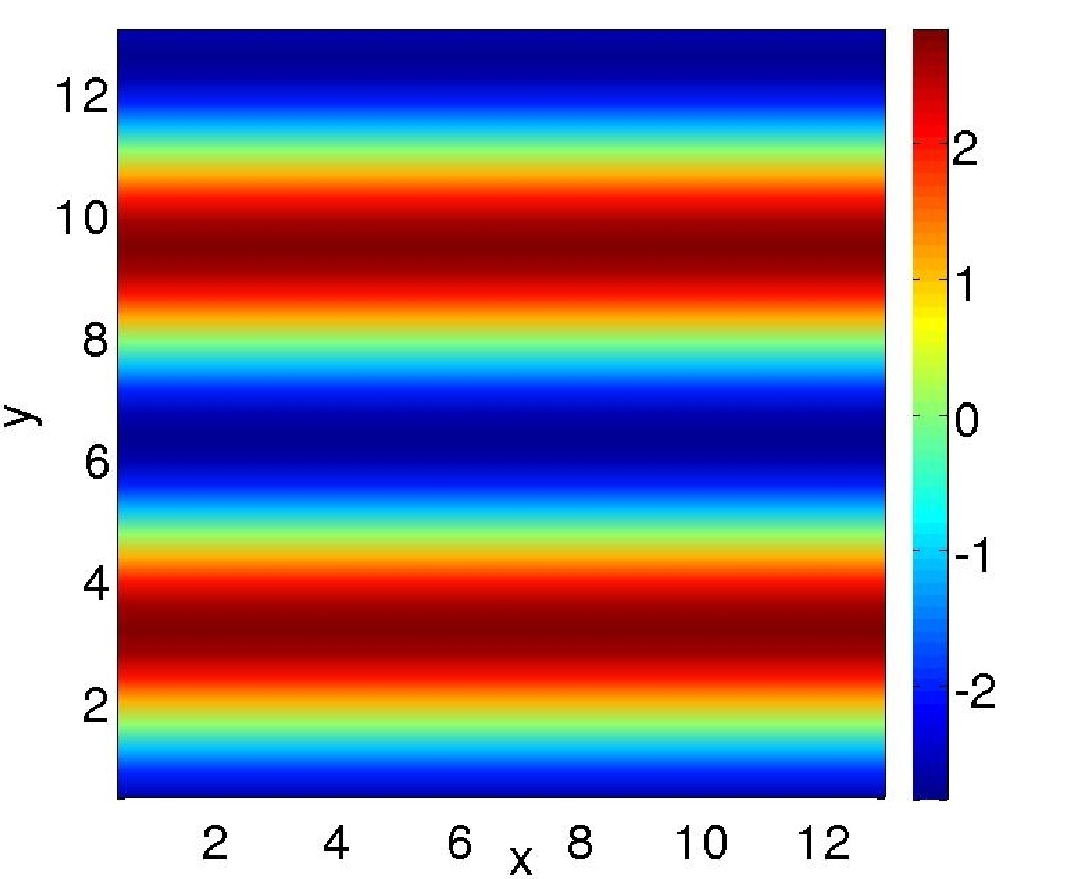}
&\includegraphics[height=0.4\linewidth]{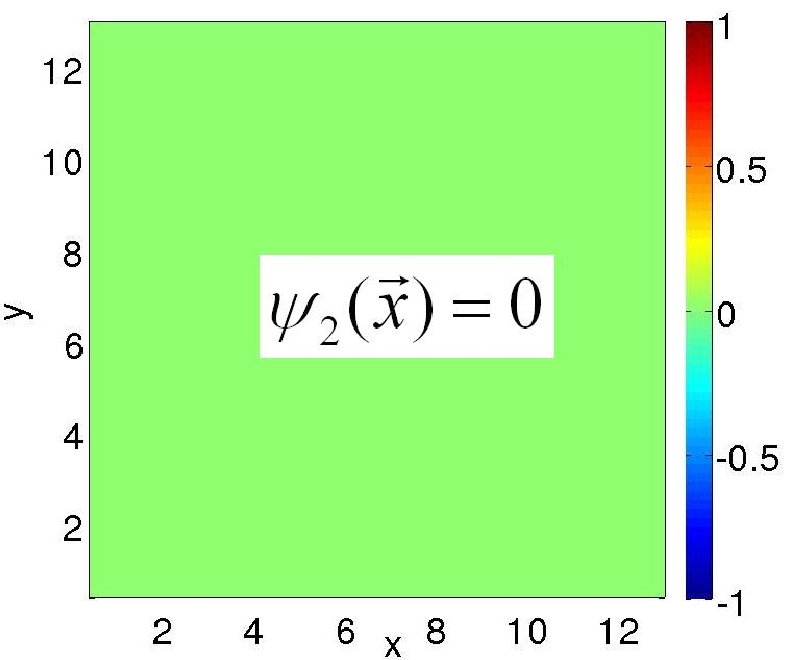}\\
\multicolumn{2}{c}{\textbf{columnar/smectic A}}\\
\includegraphics[height=0.4\linewidth]{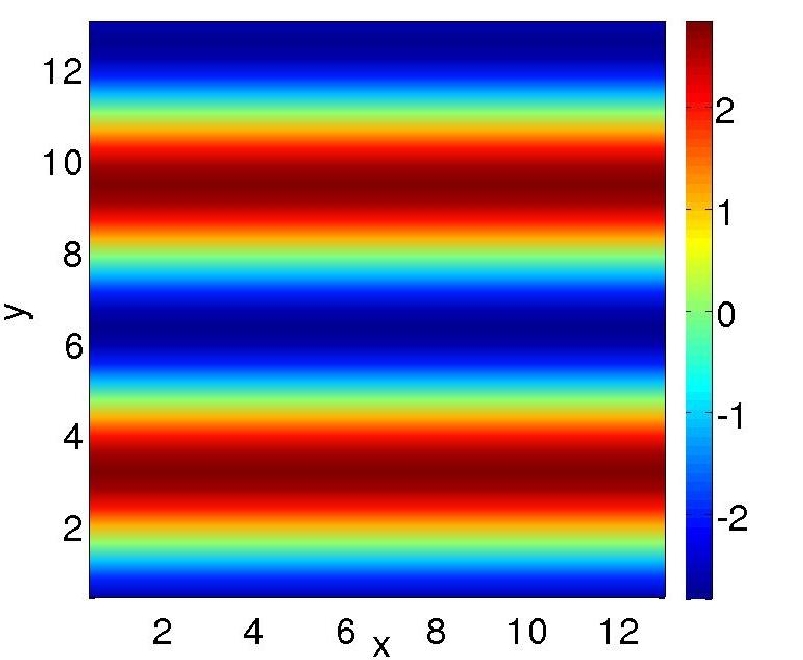}
&\includegraphics[height=0.4\linewidth]{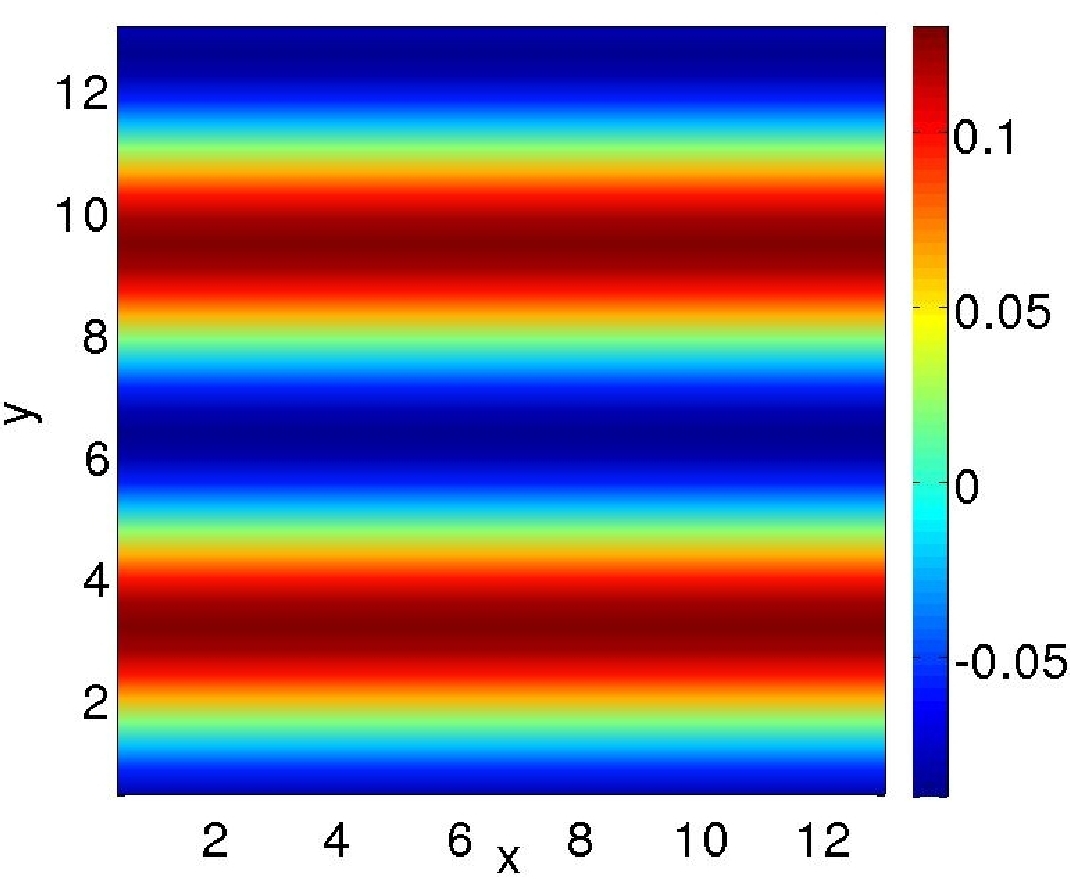}\\
\multicolumn{2}{c}{\textbf{plastic triangular crystal 1}}\\
\includegraphics[height=0.4\linewidth]{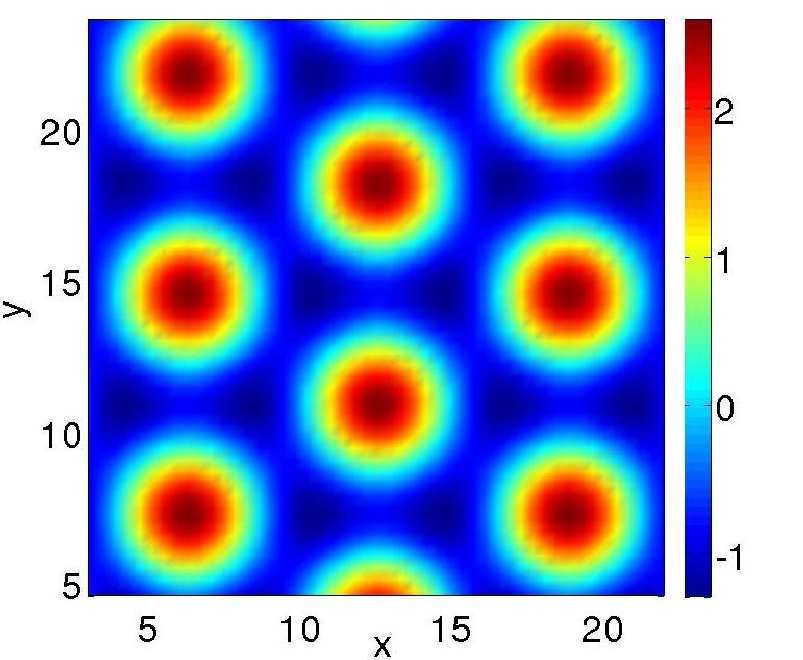}
&\includegraphics[height=0.4\linewidth]{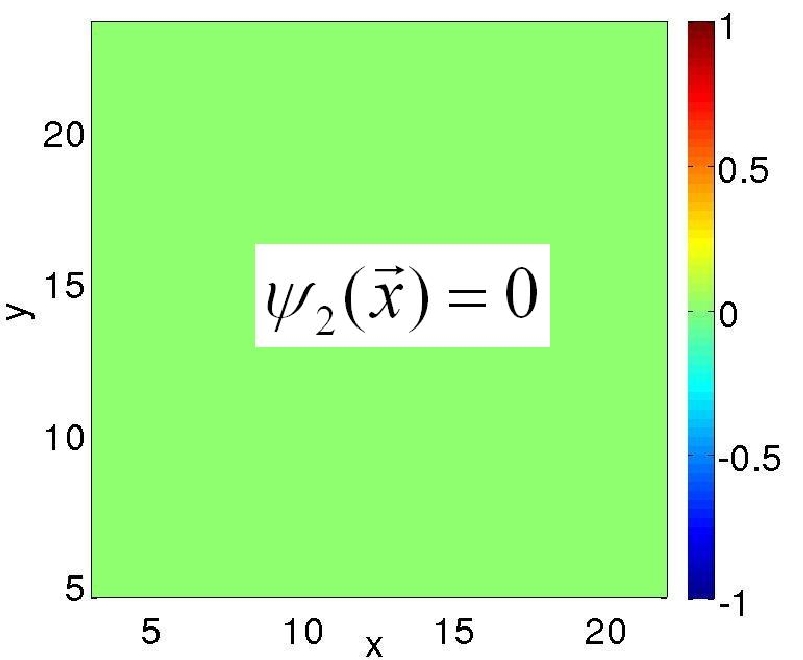}\\
$\psi_{1}(\vec{x})$
&$\psi_{2}(\vec{x})$\\
\end{tabular}
\caption{\label{fig:PB}Same as figure \ref{fig:PA}, but now for the one-mode approximation.}%
\end{figure}
\begin{figure}[ht]
\centering
\includegraphics[width=0.485\linewidth]{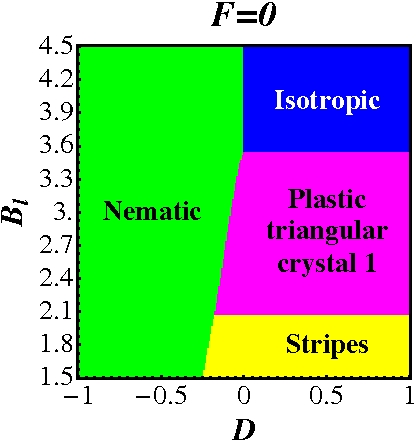}
\,
\includegraphics[width=0.485\linewidth]{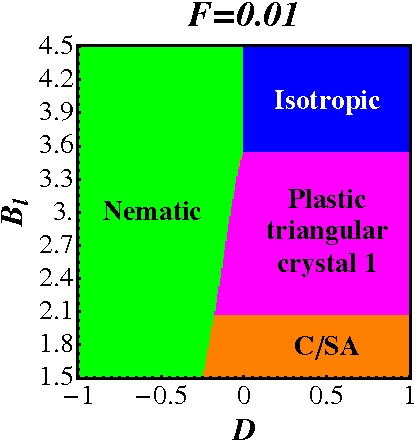}\\
\includegraphics[width=0.485\linewidth]{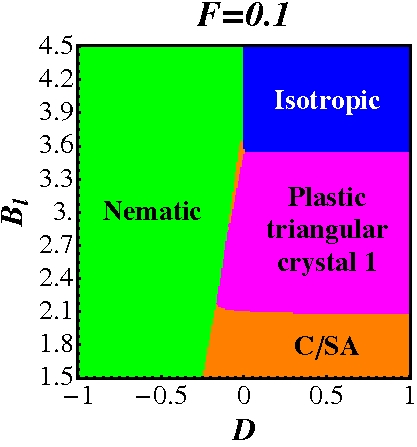}
\,
\includegraphics[width=0.485\linewidth]{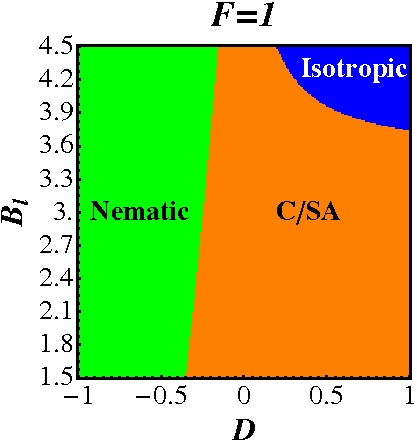}\\
\caption{\label{fig:PDO}Same as figure \ref{fig:PDA}, but now for the one-mode approximation.}%
\end{figure}
We start our investigation from the PFC model for liquid-crystalline phases, that was recently proposed in reference \cite{Loewen2010}.
It can be rescaled to the following form:
\begin{equation}
\begin{aligned}
&\mathcal{F}[\psi_{1},\psi_{2},\hat{u}_{0}] \approx \!\!
\int_{\mathbb{R}^{2}}\!\!\!\!\mathrm{d}^{2}\vec{x} \,\biggl( 
- \frac{\psi^{3}_{1}}{3} +\frac{\psi^{4}_{1}}{6} 
+(\psi_{1}-1)\frac{\psi_{1}\psi^{2}_{2}}{8} \\
&\qquad+\frac{\psi^{4}_{2}}{256} 
+B_{\mathrm{l}}\psi^{2}_{1} 
+B_{\mathrm{x}}\psi_{1}\big(2\Laplace+\Laplace^{2}\big)\psi_{1} \\[3pt] 
&\qquad+D\psi^{2}_{2} 
-E\Big(\psi_{2}\Laplace\psi_{2} +4\psi^{2}_{2}\hat{u}_{0}\!\cdot\!\Laplace\hat{u}_{0}\Big) \\
&\qquad+F\Big(\Nabla\psi_{1}\!\cdot\!\Nabla\psi_{2} +2\psi_{2} 
\!\sum^{2}_{i,j=1}\!u_{i}u_{j}\Nabla_{i}\Nabla_{j} \psi_{1}\Big)\biggr) \,.
\end{aligned}
\label{eq:FS}
\end{equation}
This functional approximates the free energy of a set of apolar uniaxial particles with the three space-dependent order-parameter
fields $\psi_{1}(\vec{x})$, $\psi_{2}(\vec{x})$, and $\hat{u}_{0}(\vec{x})$.
The field $\psi_{1}(\vec{x})$ is the reduced orientationally averaged one-particle density obtained from the general one-particle density 
$\rho(\vec{x},\hat{u})$ which corresponds to the probability density to find a particle with the orientation $\hat{u}$ at the position $\vec{x}$ as
\begin{equation}
\psi_{1}(\vec{x}) = \frac{1}{2\pi\overline{\rho}}\int_{\mathcal{S}_{1}}\!\!\!\dif\hat{u} \left(\rho(\vec{x},\hat{u})-\overline{\rho}\right) .
\end{equation}
Here, $\mathcal{S}_{1}$ is the unit circle, $\overline{\rho}$ is the mean density 
\begin{equation}
\overline{\rho} = \frac{\int\!\dif^{2}\vec{x} \int_{\mathcal{S}_{1}}\!\!\dif\hat{u} \, \rho(\vec{x},\hat{u})}{\int\!\dif^{2}\vec{x} \int_{\mathcal{S}_{1}}\!\!\dif\hat{u}} \;,
\end{equation}
and the unit vector $\hat{u}$ denotes a certain orientation in the two-dimensional space.
In comparison to the general one-particle density $\rho(\vec{x},\hat{u})$, the reduced and orientationally averaged density $\psi_{1}(\vec{x})$ only describes the space-dependent deviations of the one-particle density from the reference density $\overline{\rho}$.
Its spatial average vanishes by construction.
The orientational dependence of the free-energy is taken into account by the nematic order parameter $\psi_{2}(\vec{x})$ 
\begin{equation}
\psi_{2}(\vec{x}) = \frac{4}{\pi\overline{\rho}}\int_{\mathcal{S}_{1}}\!\!\!\dif\hat{u} 
\,\rho(\vec{x},\hat{u}) \bigg(\big(\hat{u}_{0}(\vec{x})\!\cdot\!\hat{u}\big)^{2}-\frac{1}{2}\bigg)
\end{equation}
and the nematic director $\hat{u}_{0}(\vec{x})$, which is the eigenvector associated with the largest eigenvalue of the second-order traceless nematic tensor $Q(\vec{x})$ \cite{deGennesP1995}.
The nematic director $\hat{u}_{0}(\vec{x})$ is a unit vector and can be parametrized by the orientation field $\varphi_{0}(\vec{x})$ in two spatial dimensions: $\hat{u}_{0}(\vec{x})=(\cos(\varphi_{0}(\vec{x})),\sin(\varphi_{0}(\vec{x})))$.
Also the amount of local ordering, $\psi_{2}(\vec{x})$, contributes to the free energy.
In the integrand of functional \eqref{eq:FS}, the first four terms approximate the free-energy density of an ideal rotator gas.
Also the terms $\psi^{2}_{1}(\vec{x})$ and $\psi^{2}_{2}(\vec{x})$ appear in the ideal rotator entropy,
but since there are corresponding terms in the excess free energy for anisotropic particles, the different contributions to these terms were 
combined to $B_{\mathrm{l}}\psi^{2}_{1}(\vec{x})$ and $D\psi^{2}_{2}(\vec{x})$ in Eq.\ \eqref{eq:FS}. 
Beside the mentioned polynomial terms in the order-parameter fields, also their gradients contribute to the free energy.
The amount of their contribution is controlled by the parameters $B_{\mathrm{x}}$, $E$, and $F$ in the free-energy functional.
Contributions of the gradients and curvature of the translational density field $\psi_{1}(\vec{x})$ appear in the term proportional to $B_{\mathrm{x}}$.
The curvatures of the space-dependent nematic order $\psi_{2}(\vec{x})$ and of the nematic director field $\hat{u}_{0}(\vec{x})$ are
taken into account in the term proportional to the parameter $E$.
The last term in the free-energy functional is scaled by the parameter $F$ and contains the couplings between the gradients of 
$\psi_{1}(\vec{x})$ and $\psi_{2}(\vec{x})$ as well as a sum, which couples the Hessian of $\psi_{1}(\vec{x})$ with the 
director field $\hat{u}_{0}(\vec{x})$ and with the nematic order $\psi_{2}(\vec{x})$.
In Eq.\ \eqref{eq:FS}, the components of the nematic director are denoted as $u_{i}=(\hat{u}(\vec{x}))_{i}$, for abbreviation.

It is possible to transform the functional \eqref{eq:FS} to a physically less intuitive but simpler equation by defining a new complex 
order-parameter $U(\vec{x})=\psi_{2}(\vec{x})e^{i2\varphi_{0}(\vec{x})}$.
The corresponding free-energy functional
\begin{equation}
\begin{aligned}
&\mathcal{F}[\psi_{1},U] \approx \!\!
\int_{\mathbb{R}^{2}}\!\!\!\!\mathrm{d}^{2}\vec{x} \,\biggl( 
- \frac{\psi^{3}_{1}}{3} +\frac{\psi^{4}_{1}}{6} 
+(\psi_{1}-1)\frac{\psi_{1}\psi^{2}_{2}}{8} \\
&\qquad+\frac{\psi^{4}_{2}}{256} 
+B_{\mathrm{l}}\psi^{2}_{1} 
+B_{\mathrm{x}}\psi_{1}\big(2\Laplace+\Laplace^{2}\big)\psi_{1} \\[5pt] 
&\qquad+D\,UU^{\ast} + E\,\Nabla U\!\cdot\!\Nabla U^{\ast} \\[1pt]
&\qquad+F\Big(\mathrm{Re}(U)\big(\Nabla^{2}_{1}-\Nabla^{2}_{2}\big) 
+ 2\,\mathrm{Im}(U)\Nabla_{1}\Nabla_{2}\Big)\psi_{1}\biggr)
\end{aligned}
\label{eq:FST}
\end{equation}
with $U^{\ast}(\vec{x})$ denoting the complex conjugate of $U(\vec{x})$ are helpful for more efficient numerical simulations. 
This alternative choice for the order-parameter fields is equivalent to the usage of a not 
normalized vector field instead of $\hat{u}_{0}(\vec{x})$ and the modulus $\psi_{2}(\vec{x})$, as it is done in Ginzburg-Landau theory 
\cite{deGennesP1995}.

Before we started to minimize the free-energy functional numerically, we could already derive some static properties of this functional.
Since the parameter $D$ controls the contribution of the nematic order parameter $\psi_{2}(\vec{x})$,
we expect the nematic phase to be stable for large negative values of $D$. In the opposite case, if $D$ is large enough and positive, the term $D\psi^{2}_{2}(\vec{x})+\psi^{4}_{2}(\vec{x})/256$ dominates the free energy and only phases with $\psi_{2}(\vec{x})=0$ can be stable. 
Crystalline phases with a non-vanishing nematic order might only appear in a region around $D=0$, therefore.
We also know that the difference $B_{\mathrm{l}}-B_{\mathrm{x}}$ has a big influence on the translational density field $\psi_{1}(\vec{x})$.
If the parameter $B_{\mathrm{l}}$ is large and positive, variations of the translational density field enlarge the free energy.
Similarly, gradients of the translational density field enlarge the free energy for large and negative values of $B_{\mathrm{x}}$.
Therefore, phases without any density modulations, i.e.\ the isotropic and the nematic phase, are preferred for positive values of the 
difference $B_{\mathrm{l}}-B_{\mathrm{x}}$. All other phases with a periodic translational density field are preferred for
negative values of this difference. 
Furthermore, there is a symmetry concerning the reversal of the sign of the parameter $F$ in the free-energy functional. 
From Eq.\ \eqref{eq:FS} we know that the free-energy functional is invariant under a simultaneous change of the signs of the parameter $F$ and 
the nematic order-parameter field $\psi_{2}(\vec{x})$.
Due to this symmetry, we can assume $F\geqslant 0$ in the following.

\subsection{\label{subsec:NM}Numerical minimization}
\subsubsection{\label{subsubsec:NMa}Free numerical minimization}
In order to find the stable phases in the PFC model, we minimized the free-energy functional \eqref{eq:FS} using the \textit{steepest descend method} for fixed parameters $B_{\mathrm{l}}$, $B_{\mathrm{x}}$, $D$, $E$, and $F$.
In this method, which is based on the pseudo-dynamical equations 
\begin{equation}
\begin{split}
\pdif{\psi_{1}(\vec{x},t)}{t} &= -\Fdif{\mathcal{F}[\psi_{1}(\vec{x},t),U(\vec{x},t)]}{\psi_{1}(\vec{x},t)}+\lambda(t) \;, \\
\pdif{U(\vec{x},t)}{t} &= -\Fdif{\mathcal{F}[\psi_{1}(\vec{x},t),U(\vec{x},t)]}{U^{\ast}(\vec{x},t)} \;,
\end{split}
\label{eq:umot}
\end{equation}
the system evolves towards a local minimum. 
Here, $\lambda$ is a Lagrange multiplier that guarantees that the spatial average of the field $\psi_{1}(\vec{x})$ is zero.
The two equations were discretized using a finite-difference scheme and solved on a grid with $32\times 32$ cells.
We varied the length of these cells to minimize the free-energy functional also with respect to the lattice constant of the periodic phases 
and used a set of different phases as initial conditions in order to find the global minimum.

\subsubsection{\label{subsubsec:NMb}One-mode approximation}
As a semi-analytical approach the \textit{one-mode approximation} consists of periodic approximations for the order-parameter fields $\psi_{1}(\vec{x})$, $\psi_{2}(\vec{x})$, and $\varphi_{0}(\vec{x})$ and reduces the PFC model to the lowest Fourier modes.
We used the following parametrizations with the minimization parameters $A_{1}$, $B_{0}$, $B_{1}$, and $k$:
\begin{itemize}
\item isotropic phase: 
\begin{equation}
\begin{split}
\psi_{1}(\vec{x})&=0 \;, \\
\psi_{2}(\vec{x})&=0 \;,
\end{split}
\label{eq:I}
\end{equation}
\item nematic phase:
\begin{equation}
\begin{split}
\psi_{1}(\vec{x})&=0 \;, \\
\psi_{2}(\vec{x})&=B_{0} \;,
\end{split}
\label{eq:N}
\end{equation}
\item stripe phase:
\begin{equation}
\begin{split}
\psi_{1}(\vec{x})&=A_{1}\cos(ky) \;, \\
\psi_{2}(\vec{x})&=0 \;,
\end{split}
\label{eq:S}
\end{equation}
\item columnar phase:
\begin{equation}
\begin{split}
\psi_{1}(\vec{x})&=A_{1}\cos(ky)\;, \\
\psi_{2}(\vec{x})&=B_{0} + B_{1}\cos(ky)\;, \\
\varphi_{0}(\vec{x})&=0 \;,
\end{split}
\label{eq:C}
\end{equation}
\item smectic A phase:
\begin{equation}
\begin{split}
\psi_{1}(\vec{x})&=A_{1}\cos(ky)\;, \\
\psi_{2}(\vec{x})&=B_{0} + B_{1}\cos(ky)\;, \\
\varphi_{0}(\vec{x})&=\frac{\pi}{2} \;,
\end{split}
\label{eq:SA}
\end{equation}
\item triangular crystalline phase:
\begin{equation}
\begin{split}
\psi_{1}(\vec{x})&= A_{1}\biggl(\!\cos\Big(\frac{\sqrt{3}}{2}k x\Big)\cos\Big(\frac{k}{2}y\Big) \\
&\qquad\quad\! -\frac{\cos(ky)}{2}\!\biggr) \,, \\
\psi_{2}(\vec{x})&= B_{0} + B_{1}\biggl(\!\cos\Big(\frac{\sqrt{3}}{2}k x\Big)\cos\Big(\frac{k}{2}y\Big) \\
&\qquad\qquad\quad\; -\frac{\cos(ky)}{2}\!\biggr) \,, \\
\varphi_{0}(\vec{x})&=\phi_{0} \;,
\end{split}
\label{eq:TK}
\end{equation}
\item square crystalline phase:
\begin{equation}
\begin{split}
\psi_{1}(\vec{x})&=A_{1}\big(\cos(kx)+\cos(ky)\big)\;, \\
\psi_{2}(\vec{x})&=B_{0} + B_{1}\cos(kx)\cos(ky)\;, \\
\varphi_{0}(\vec{x})&=\phi_{0} \;.
\end{split}
\label{eq:QK}
\end{equation}
\end{itemize}
Here, $x$ and $y$ are the first and the second Cartesian coordinate, respectively. 
The constant angle $\phi_{0}$ can be set to zero, because the free energy does not depend on it.
Further, due to equivalent free energies, we do not need to distinguish between the columnar phase and the smectic A phase and call them columnar/smectic A phase in the following.
Note, that there is no additive offset term $A_{0}$ for the density variations $\psi_{1}(\vec{x})$.
The minimization of the free energy in the context of the one-mode approximation was performed for fixed parameters $B_{\mathrm{l}}$, $B_{\mathrm{x}}$, $D$, $E$, and $F$.
It is possible to minimize the free energy for the nematic phase and for all phases with a vanishing nematic order $\psi_{2}(\vec{x})$ 
analytically.
The more complicated free energies of the remaining phases were minimized numerically.

\section{\label{sec:NR}Numerical results}
\subsection{\label{subsec:NRa}Free numerical minimization}
Apart from the fully isotropic phase ($\psi_{1}(\vec{x})=0$, $\psi_{2}(\vec{x})=0$), which appears for $B_{\mathrm{l}}>B_{\mathrm{x}}$ and $D>0$, several other phases were found to minimize the free-energy (see figure \ref{fig:PA}). As expected, for negative and large $D$, a nematic phase was found. In the columnar/smectic A phase, the system has positional ordering in one direction while it is isotropic perpendicular to this direction.
The $\psi_{2}(\vec{x})$ field has a similar profile to the $\psi_{1}(\vec{x})$ field with maxima of these two fields at the same positions. 
The director field $\hat{u}_{0}(\vec{x})$ is oriented parallel to the gradient of the field $\psi_{1}(\vec{x})$, $\Nabla\psi_{1}(\vec{x})$, in the regions with negative $\Nabla^{2}_{1}\psi_{1}(\vec{x})$, while for positive $\Nabla^{2}_{1}\psi_{1}(\vec{x})$ it is perpendicular to the gradient.
A similar flipping of the orientational field from perpendicular to parallel to the stripe direction was identified as transverse intralayer order
in the three-dimensional smectic A phase of hard spherocylinders \cite{vanRoijBMF1995}.

Four plastic crystals with different symmetries were found.
The first two phases are plastic triangular crystals with a vanishing and a non-vanishing nematic order parameter. 
Here, the first crystal with $\psi_{2}(\vec{x})=0$ is a special degenerate case of the second. 
The third plastic crystalline structure involves a honeycomb lattice.
As a fourth case, there is a plastic crystal with square symmetry.
For all plastic crystals, $\psi_{2}(\vec{x})$ vanishes both at the maxima and minima of $\psi_{1}(\vec{x})$.

The director fields $\hat{u}_{0}(\vec{x})$ for the different plastic crystalline phases exhibit quite different topologies.
While the director field is not defined for the plastic triangular crystal 1 with a vanishing field $\psi_{2}(\vec{x})$, 
it possesses in general topological defects at positions where the field $\psi_{2}(\vec{x})$ vanishes. This guarantees that there
is a finite core energy of the topological defect \cite{TasinkevychSPTdG2002}.
The topological defects form another lattice with more lattice points than that given by the maxima of the field $\psi_{1}(\vec{x})$
since there are additional \textit{interstitial} topological defects at the minima of $\psi_{1}(\vec{x})$.
The lattices of topological defects are schematically schown in figure \ref{fig:D}.
For the plastic triangular crystal 2 and for the plastic honeycomb crystal, the associated defect crystal is triangular albeit with a lattice constant that is a factor of $1/\sqrt{3}$ smaller than the original one.
Likewise, for the plastic square crystal, the defect lattice is a square lattice with a lattice constant reduced by a factor $1/\sqrt{2}$.
Topological defects in liquid crystals can be classified according to the winding number of their director field \cite{deGennesP1995,Mermin1979}.
In our case, three types of point defects occur: vortices with the topological winding number $m=1$, sources/sinks with $m=1$, and hyperbolic points with $m=-1$.
Furthermore, we found disclination line defects with $m=-1/2$.
Vortices and disclination lines occur in the plastic triangular crystal 2, which is schematically drawn in the first plot in figure \ref{fig:D}.
In the plastic honeycomb crystal, disclination lines arise together with sources/sinks (second plot in figure \ref{fig:D}), while vortices and hyperbolic points are found in the plastic square crystal (last plot in figure \ref{fig:D}).
The sum of the topological winding numbers of all topological defects in a unit cell vanishes for all plastic crystals.

The director fields of all crystalline phases that we found are periodic and vanish when they are averaged in space.
Therefore, these crystals are identified as being plastic.
Orientationally ordered crystalline phases were not found in the parameter range we explored.

Parameters for which the mentioned phases are stable follow from the phase diagrams in figure \ref{fig:PDA}.
We chose $1.5\leqslant B_{\mathrm{l}}\leqslant 4.5$, $B_{\mathrm{x}}=3.5$, $-1\leqslant D\leqslant 1$, $E=0.1$, and $F=0,0.01,0.1,1$ for these phase diagrams. We chose $B_{\mathrm{x}}$ constant and varied $B_{\mathrm{l}}$ so that in the absence of the orientational degrees of freedom we obtain all the phases of the original PFC model. The parameter $D$ was varied from $-1$ because for this value the energy of the nematic phase was significantly smaller than the energy of any other phase of the original PFC model. The parameters $E$ and $F$ were selected from the regions where we found the richest phase diagrams.
For $F=0$ we have a degenerate case, where $\psi_{2}(\vec{x})=0$ for $D>0$ and $E>0$.
In the latter case we observed a stripe phase and the plastic triangular crystal 1.
These phases are replaced by the columnar/smectic A phase and by the plastic triangular crystal 2 if $F$ becomes positive.
The richest phase diagram with six different phases was obtained for $F=1$.
The phase transition between the isotropic and the nematic phase turned out to be continuous, while all other phase transitions are discontinuous.
This result agrees with the fact that the PFC model reduces to the Landau-de Gennes model for the isotropic-nematic phase transition, which describes this phase transition as continuous. On the other hand, symmetries are broken for the remaining phase transitions so that they are discontinuous.

\subsection{\label{subsec:NRb}One-mode approximation}
For the global minimization of the free energy we used a random search routine in the four-dimensional parameter space combined with a local minimization by a Newton method.
Figure \ref{fig:PB} shows the order-parameter fields $\psi_{1}(\vec{x})$ and $\psi_{2}(\vec{x})$ for different phases that are realized for particular combinations of the parameters of the PFC model in the one-mode approximation.
To calculate phase diagrams that correspond to the phase diagrams presented in Subsec.\ \ref{subsec:NRa}, we fixed the parameters $B_{\mathrm{x}}$, $E$, and $F$ to the same values as in the previous subsection, varied the remaining parameters of the rescaled free-energy functional \eqref{eq:FS} over the intervals $-1\leqslant D\leqslant 1$ and $1.5\leqslant B_{\mathrm{l}}\leqslant 4.5$, and minimized the free-energy for each one-mode approach and for each parameter combination.
The resulting phase diagrams are shown in figure \ref{fig:PDO}.
A view at the behavior of the order parameters near to phase transitions confirmed the results of the free numerical minimization regarding the order of the phase transitions.

\subsection{\label{subsec:NRc}Comparison and discussion}
In comparison to the full numerical minimization of the free-energy functional, the one-mode approximation appears to provide an approximative but much faster method to determine the phase diagram for the PFC model.
A speed-up factor of 100 can easily be achieved. This allows the fast calculation of phase diagrams for various parameter combinations as well as single phase diagrams with a rather high resolution, like those that are shown in figure \ref{fig:PDO}.
The drawback, however, is that the one-mode approximation also gives rise to deviations of the phase diagrams from the exact ones in figure \ref{fig:PDA}: there is a lack of some crystalline phases for $F=1$.
The absence of these crystalline phases is based on an improper consideration of the orientation field $\varphi_{0}(\vec{x})$ in the 
one-mode approximation and the last term in the free-energy functional \eqref{eq:FS} which is relevant for non-vanishing values of the parameter $F$.
Therefore, the plastic square crystal as well as the plastic honeycomb crystal cannot be observed in the one-mode approximation.
In addition, the plastic triangular crystal 1 with $\psi_{2}(\vec{x})=0$ appears for $F>0$ instead of the plastic triangular crystal 2.
This is not surprising since we used a constant angle $\phi_{0}$ for the orientation field in our ansatz for the one-mode approximation to 
parametrize plastic crystals without a global orientation.
The constant angle $\phi_{0}$ is not a good approximation for the actual orientation field with a vanishing mean value so that the minimum of the 
free energy is reached for $\psi_{2}(\vec{x})=0$.
A further difference between the phase diagrams in figure \ref{fig:PDO} and those in figure \ref{fig:PDA} is an island of the columnar/smectic A phase near to the lower end of the phase transition line between the isotropic and the nematic phase for $F=0.1$.

To check that the mentioned disadvantages of the phase diagrams for the one-mode approximation really arise from the improper approximation 
of the orientation field $\varphi_{0}(\vec{x})$, we performed a Fourier analysis of the numerical results that are shown in figure \ref{fig:PA}.
This Fourier analysis exhibited that the first Fourier mode is dominant. 
Only for the plastic square crystal and for the plastic honeycomb crystal there appears a contribution of the second mode that might be relevant.

Despite the discussed disadvantages of the one-mode approximation, it proved to be an useful method in order to calculate phase diagrams for a given PFC model with low computational effort.
The one-mode approximation is also useful to explore the phase diagram for suitable parameters or for a large number of parameter sets and to find interesting regions in a high-dimensional parameter space that are worth to be investigated more precise by a much more expensive direct minimization 
of the free-energy functional.

\section{\label{sec:conclusions}Conclusions}
In conclusion, the phase-field crystal model for liquid crystals was solved numerically
in two spatial dimensions and different stable phases were found in the parameter space of the model.
These include isotropic, nematic, columnar, smectic A, and plastic crystalline phases.
The latter can possess a periodic triangular, square or honeycomb structure with a complex orientation field.
The phase-diagrams were obtained by a numerical minimization of
the free-energy functional but are - in most but not all cases - qualitatively in line with much simpler one-mode approximations
for the order parameter fields. We hope that this work will inspire more simulations and experiments
to observe the unexpected predicted phases (like the square and honeycomb crystals) in two dimensions
\cite{WatanabeT2007,Vink2007,TavaresHTdG2009,TothDY2002,LopezLRPC2010}. One can think about the realization of molecular liquid crystals as well as
concentrated solutions of anisotropic colloidal particles \cite{VroegeL1992,vanderBeekDWVL2008}
or anisotropic mesoscopic dust particles in a plasma \cite{AnnaratoneEtAl2001,IvlevKKAMY2003}.
In particular, it would be challenging to explore the orientation field and the associated defect lattice in plastic crystals either by simulations 
\cite{MarechalD2008} or by experiments \cite{Demiroers2010,GerbodeAOLEC2010}.

An important further step is to calculate the full liquid-crystalline phase diagram for a given interparticle potential
as a function of the real thermodynamic variables, namely temperature and number density.
This phase diagram is known from computer simulations and from theory for hard
spherocylinders \cite{BolhuisF1997,GrafLS1997}, for hard ellipsoids \cite{FrenkelM1985}, for the Gay-Berne \cite{deMiguelMdRBA1996,KondratBH2010}, and for Yukawa segment models \cite{GrafL1999}. In order to do this one has to map the system at given temperature and
density onto the parameter space. This mapping needs the full direct correlation function of the isotropic phase as an input.
The latter can be obtained either by simulation and liquid-integral equations for anisotropic systems \cite{WeyerichDACK1990} or by
fundamental-measure-theory for anisotropic hard particles \cite{HansenGoosM2009}.

In future work the model can be applied and solved for more complicated situations
such as interfaces between two coexisting phases. While the isotropic-nematic interface has been studied
by theory, simulation and experiment \cite{McDonaldAS2001,vanderBeekEtAl2006}, it would be in particular interesting
to get structural information about solid-fluid and solid-solid interfaces. Then the interfacial tension is anisotropic
and the degree of anisotropy can be extracted from the phase-field-crystal approach.
The structure of topological defects would be another playground where the present approach could be applied to
\cite{TothDY2002,LiuM1997}.

Moreover, the extension
towards dynamics is straightforward \cite{Loewen2010} and the numerical implementation can in principle be done.
We expect a wealth of different dynamical growth effects \cite{BechhoeferLT1991} and novel steady states in
systems driven by external \cite{HaertelBL2010} and internal \cite{TonerTR2005,WensinkL2008,vanTeeffelenL2008}
forces. Finally a big challenge is to implement the functional in three spatial dimensions where there are more
coupling parameters \cite{WittkowskiLB2010}.

\begin{acknowledgments}
We thank H. R. Brand, H. Emmerich, U. Zimmermann, M. Oettel, and M. Marechal for helpful discussions. 
This work has been supported by the DFG through the DFG priority program SPP 1296. 
\end{acknowledgments}

\bibliography{References}
\bibliographystyle{rsc}
\end{document}